# Quantum antireflection temporal coatings: quantum state frequency shifting and inhibited thermal noise amplification

Iñigo Liberal[1], J. Enrique Vázquez-Lozano[1] and Victor Pacheco-Peña[2]

[1]Department of Electrical, Electronic and Communications Engineering, Institute of Smart Cities (ISC), Public University of Navarre (UPNA), 31006 Pamplona, Spain

[2]School of Mathematics, Statistics and Physics, Newcastle University, Newcastle Upon Tyne, NE1 7RU, UK

**Abstract –** We investigate the quantum optical response of antireflection temporal coatings, i.e., matching temporal layers that suppress the generation of backward waves in temporal boundaries. Our results reveal that quantum antireflection temporal coatings are characterized for inducing a frequency shift of the quantum state, while preserving all photon statistics intact. Thus, they might find application for fast quantum frequency shifting in photonic quantum networks. The quantum theory also provides additional insight on their classical mode of operation, clarifying which quantities are preserved through the temporal boundary. Finally, we show that quantum antireflection temporal coatings allow for fast temporal switching without the amplification of thermal fields.

1. Introduction

Temporal metamaterials (materials with a designed temporal variation of their constitutive parameters) provide an additional and fundamentally different degree of freedom in engineering light-matter interactions [1–5]. The elementary constituent of temporal metamaterials is the temporal boundary, i.e., the rapid change in time of the material parameters of the system [6,7]. The main signature of a temporal boundary is the generation of forward and backward waves (temporal equivalent of transmitted and reflected signal at a spatial boundary between two media), shifted in frequency, but preserving the wavenumber in order to enforce the continuity of the ***D*** and ***B*** fields [6–9].

Since the generation of a backward wave can be undesired in many practical scenarios, antireflection temporal coatings have recently emerged as a design strategy to achieve reflection-less temporal boundaries [10]. Inspired by the design of spatial matching layers [11], temporal slabs can be tailored such that destructive interference inhibits the generation of a backward wave. At the same time, given that spatial interfaces and temporal boundaries have different properties, spatial and temporal matching layers also present different features, as discussed in [12–14]. The extension to multilayered temporal sequences enables a finer control over the spectral response of antireflection temporal coatings. Several approaches have pushed forward this concept, leveraging strategies from advanced filter design [15], granting access to higher-order functions [14], and enabling ultrawideband pulse operation [16]. Further extensions to a very large number of temporal slabs can be effectively described though effective medium theories [17–19] and/or as photonic time crystals with nontrivial topological properties [20–22].

The quantum optical response of temporal media is also of interest, resulting in squeezing transformations [23], photon production from superluminal boundaries [24] and temporal beamsplitters [25]. Temporal metamaterials also allow for controlling quantum radiative transitions in localized [26] and free-electron emitters [27].

In this work we investigate the quantum response of antireflection temporal coatings, hereafter referred to as quantum antireflection temporal coatings. As we will show, the quantum theory provides additional physical insight on their classical mode of operation. Moreover, it points towards additional potential applications of antireflection temporal coatings such as quantum frequency shifting and fast switching while inhibiting the amplification of thermal fields.

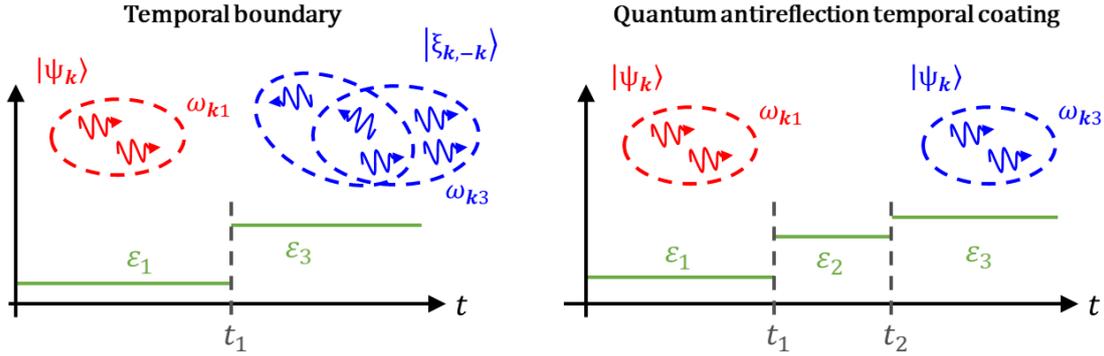

*Fig. 1. Quantum state transformations in temporal boundaries. (Left) Sketch of the change of a quantum state through a temporal boundary, including the generation of forward and backward photons, with nontrivial quantum correlations between them. (Right) Quantum antireflection temporal coatings preserve the quantum state, while shifting it in frequency.*

## 2. Quantum theory of antireflection temporal coatings

The general structure of an antireflection temporal coating is schematically depicted in Fig. 1. It basically consists of two semi-infinite temporal slabs with permittivity $\varepsilon_1$ and $\varepsilon_3$, linked by a temporal matching layer with permittivity $\varepsilon_2$ and duration $\tau = t_2 - t_1$. All temporal layers are assumed to be nonmagnetic ($\mu_1 = \mu_2 = \mu_3 = 1$).

On each temporal slab, the Hamiltonian of the system is $\hat{H}_n = \sum_{\boldsymbol{k}} \hbar \omega_{kn} \hat{a}_{\boldsymbol{k}n}^\dagger \hat{a}_{\boldsymbol{k}n} + 1/2$, with $n = 1,2,3$, representing a continuum of photonic modes with wavevector $\boldsymbol{k}$, frequency $\omega_{kn} = |\boldsymbol{k}|c/\sqrt{\varepsilon_n}$ and destruction operator $\hat{a}_{\boldsymbol{k}n}$. The electric field operator is given by [28]

$$\hat{\boldsymbol{E}}_{\boldsymbol{k}n} = i \sqrt{\frac{\hbar \omega_{kn}}{2\varepsilon_0 \varepsilon_n V}} \, \boldsymbol{e}_{\boldsymbol{k}n} \, e^{i\boldsymbol{k}\cdot\boldsymbol{r}} \, \hat{a}_{\boldsymbol{k}n} + h.c. \qquad (1)$$

where $V$ is the quantization volume and $\boldsymbol{e}_{\boldsymbol{k}n}$ is the unit polarization vector. Imposing the continuity of the $\hat{\boldsymbol{D}}_{\boldsymbol{k}n}$ and $\hat{\boldsymbol{B}}_{\boldsymbol{k}n}$ operators across the temporal boundaries leads to the following transformation rule for the photonic operators

$$\hat{a}_{\boldsymbol{k}3} = \left(\cosh(s_{21})\cosh(s_{32}) + e^{i2\omega_{k2}\tau}\sinh(s_{21})\sinh(s_{32})\right)\hat{a}_{\boldsymbol{k}1} \qquad (2)$$

$$-\left(\sinh(s_{21})\cosh(s_{32}) + e^{i2\omega_{k2}\tau}\sinh(s_{32})\cosh(s_{21})\right)\hat{a}^\dagger_{-\mathbf{k}1}$$

where $s_{mn} = \ln\sqrt{Z_m/Z_n}$ is the squeezing parameter characterizing the temporal boundary between the m$^{th}$ and the n$^{th}$ temporal slabs and $Z_m = 1/\sqrt{\varepsilon_m}$ is the medium impedance in the m$^{th}$ temporal slab. The transformation rule given by Eq. (2) has two major characteristics: (i) the combination of forward $\mathbf{k}$ and backward $-\mathbf{k}$ modes, leading to nontrivial photon correlations between them. (ii) The mixing of creation and annihilation operators, resulting in photon production and/or absorption effects. In general, these two characteristics induce a change in the quantum state of the system, as schematically depicted in Fig. 1.

For the particular case in which the duration of the temporal slab equals a quarter of the period of the frequency of the mode with wavevector $\mathbf{k}$ inside it, i.e., $2\omega_{k2}\tau = \pi$ or $\tau = T_{k2}/4$, the operator transformation rule (2) reduces to

$$\hat{a}_{\mathbf{k}3} = \cosh(s_{32} - s_{21})\,\hat{a}_{\mathbf{k}1} - \sinh(s_{32} - s_{21})\hat{a}^\dagger_{-\mathbf{k}1} \tag{3}$$

In other words, a $T_{k2}/4$ temporal slab results in a two-mode squeezing transformation, where the effective squeezing parameter is the subtraction of the squeezing parameter for the first and second temporal boundaries composing the temporal slab. Equivalently, it can be stated that the impact of a $T_{k2}/4$ temporal slab is that of a single temporal boundary, with effective squeezing parameter $s_{32} - s_{21}$.

In view of Eq. (3), it is clear that a case of particular interest will be $s_{32} - s_{21} = 0$, i.e., when the squeezing parameters of the first and second temporal boundaries exactly cancel out. Since $s_{mn} = \ln\sqrt{Z_m/Z_n}$, it can be readily checked that this condition is equivalent to $Z_2 = \sqrt{Z_1 Z_3}$. That is, the impedance of the matching layer should equal the geometric mean of the initial and final medium impedances. This condition is exactly equivalent to that of a classical $T/4$ antireflection temporal coating [10], or a traditional $\lambda/4$ impedance matching network in the spatial domain [11]. In such a case, the operator transformation rule for an antireflection temporal coating simply reduces to

$$\hat{a}_{\mathbf{k}3} = \hat{a}_{\mathbf{k}1} \tag{4}$$

This simple relation between the operators before and after the temporal slab has a clear physical meaning: Quantum states are projected with no changes in their photon statistics, preserving all their quantum correlations. At the same time, the frequency of the photonic modes is shifted from $\omega_{\mathbf{k}1} = |\mathbf{k}|c/\sqrt{\varepsilon_1}$ to $\omega_{\mathbf{k}3} = |\mathbf{k}|c/\sqrt{\varepsilon_3}$ (after applying the second temporal boundary at $t = t_2$). Therefore, an antireflection temporal coating allows for a quantum state to be perfectly transferred through a temporal sequence, while shifting it in frequency.

Since high quality quantum sources are typically restricted to a set of specific wavelengths, quantum frequency conversion is a major technological challenge in the current development of quantum technologies [29–34]. In particular, frequency shifting nonclassical light states is required to operate at frequencies where propagation losses are small, to fully exploit the bandwidth, to develop tunable nonclassical light sources, or to connect with other elements of the photonic network such as a quantum memory. Antireflection temporal coatings might be of interest as they provide a fast frequency shifting procedure, the process only taking the duration of the matching layer, i.e., a quarter of the period within it.

### 3. Revisiting the classical case

Next, we use the quantum formalism to revisit and bring new insights into the classical response of antireflection temporal coatings [10]. To represent a classical wave, we set the initial state of the system as a coherent state $|\psi_{in}\rangle = \widehat{D}_{\mathbf{k1}}(\alpha)|0\rangle$, where $\widehat{D}_{\mathbf{k1}}(\alpha) = \exp\left(\alpha \hat{a}^\dagger_{\mathbf{k1}} - \alpha^* \hat{a}_{\mathbf{k1}}\right)$ is the displacement operator with complex amplitude $\alpha = \alpha' + i\alpha''$ [28]. The configuration is schematically depicted in Fig. 2.

As anticipated, the major feature of a quantum antireflection temporal coating is that it preserves the quantum state and all its photon statistics. That is to say, the average number of photons $\langle n_{\mathbf{k1}}\rangle = \langle n_{\mathbf{k3}}\rangle = |\alpha|^2$, the amplitude $\langle X_{\mathbf{k1}} + iY_{\mathbf{k1}}\rangle = \alpha = \langle X_{\mathbf{k3}} + iY_{\mathbf{k3}}\rangle$, and the variances $\Delta X^2_{\mathbf{k1}} = \Delta Y^2_{\mathbf{k1}} = \frac{1}{4} = \Delta X^2_{\mathbf{k3}} = \Delta Y^2_{\mathbf{k3}}$ of the quadrature operators, $X_{\mathbf{k}n} = \left(\hat{a}^\dagger_{\mathbf{k}n} + \hat{a}_{\mathbf{k1}}\right)/\sqrt{2}$ and $Y_{\mathbf{k}n} = i\left(\hat{a}^\dagger_{\mathbf{k}n} - \hat{a}_{\mathbf{k1}}\right)/\sqrt{2}$, as well as any other photon statistic will remain those of the initial coherent state (see Fig. 3a).

At the same time, preserving photon statistics does not imply that all physical quantities will remain invariant. Up and down frequency conversion takes place according to the ratio: $\omega_{\mathbf{k3}}/\omega_{\mathbf{k1}} = \sqrt{\varepsilon_1/\varepsilon_3}$. Similarly, the energy in the system scales from $\langle \widehat{H}_{\mathbf{k1}}\rangle = \hbar\omega_{\mathbf{k1}}|\alpha|^2$ to $\langle \widehat{H}_{\mathbf{k3}}\rangle = \hbar\omega_{\mathbf{k3}}|\alpha|^2$, directly following the frequency change (see Fig. 2b). Interestingly, a quantum antireflection coating modifies the energy of the electromagnetic field via a pure frequency shifting, without involving any absorption and/or photon production process. Finally, as shown in Fig. 2c, the amplitude of the electric field changes following a $\langle \hat{E}_{\mathbf{k3}}\rangle/\langle \hat{E}_{\mathbf{k1}}\rangle = (\varepsilon_1/\varepsilon_3)^{3/4}$ scaling factor, in accordance to the $\sqrt{\omega_{\mathbf{k}n}/\varepsilon_n}$ prefactor of the electric field operator in Eq. (1).

These scaling factors are consistent with recent discussions on the operation principle of classical antireflection temporal coatings, and the extent of their analogy with spatial matching layers [12]. In fact, Fig. 2c also includes a numerical calculation for the classical prediction of the change in the energy and electric field amplitude, showing an excellent agreement. We remark that, for most temporal boundaries, the quantum and classical predictions of the change in the energy do not match due to the amplification of the quantum noise in coherent states. However, this effect is inhibited in quantum antireflection temporal coatings.

The quantum theory provides a more complete picture of the operational principle of antireflection temporal coatings: quantum states are perfectly transferred through them, with no changes in their photon statistics. All observed differences are justified by the fact that, for a different background medium, each quanta of light has a different energy, a different electric field strength, etc.

Finally, we emphasize that, from the quantum theory perspective, the impact of an unmatched temporal boundary goes beyond the mere generation of a backward wave. If we were to switch between media 1 and 3 without the temporal matching layer (i.e., using a simple temporal boundary) the operator transformation rule would be that of a single temporal boundary: $\hat{a}_{\mathbf{k3}} = \cosh(s_{31})\hat{a}_{\mathbf{k1}} - \sinh(s_{31})\hat{a}^\dagger_{-\mathbf{k1}}$. Consequently, there would be changes in the output photon statistics for both the forward and backward waves, including the amplitude of the quadrature operators $\langle X_{k3} + iY_{k3}\rangle = \alpha_{k3} = \cosh(s_{31})\,\alpha$ and $\langle X_{-k3} + iY_{-k3}\rangle = \alpha_{-k3} = -\sinh(s)\,\alpha^*$, the average number of photons, $\langle n_{k3}\rangle = |\alpha_{sk3}|^2 + \sinh^2(s_{31})$ and $\langle n_{-k3}\rangle = |\alpha_{-k3}|^2 + \sinh^2(s_{31})$, and the variances $\Delta X_{k3}^2 = \Delta Y_{k3}^2 = \Delta X_{-k3}^2 = \Delta Y_{-k3}^2 = \frac{1}{4}\cosh(2s)$, as represented in Fig. 3b.

In addition, the existence of a backward wave imposes nontrivial photon correlations that cannot be appreciated in the classical case. To illustrate this point, we can rewrite the photonic operators in a basis formed by symmetric $\hat{a}_{s\mathbf{k}} = (\hat{a}_{\mathbf{k}} + \hat{a}_{-\mathbf{k}})/\sqrt{2}$ and antisymmetric $\hat{a}_{a\mathbf{k}} = (\hat{a}_{\mathbf{k}} + \hat{a}_{-\mathbf{k}})/\sqrt{2}$ operators, which makes more evident the squeezing nature of the transformation. In this basis, the amplitude of the quadrature operator changes from $\langle X_{sk1} + iY_{sk1}\rangle = \alpha/\sqrt{2} = \langle X_{ak1} + iY_{ak1}\rangle$ to $\langle X_{sk3} + iY_{sk3}\rangle = \alpha_{a3} = \frac{1}{\sqrt{2}}(e^{-s_{31}}\alpha' + i\,e^{s_{31}}\alpha'')$ and to $\langle X_{ak3} + iY_{ak3}\rangle = \alpha_{a3} = \frac{1}{\sqrt{2}}(e^{s_{31}}\alpha' + i\,e^{-s_{31}}\alpha'')$, as shown in Fig. 3. Similarly, the number of photons changes from $\langle n_{sk1}\rangle = \langle n_{ak1}\rangle = |\alpha|^2/2$ to $\langle n_{sk3}\rangle = |\alpha_{s3}|^2 + \sinh^2(s_{31})$ and $\langle n_{ak3}\rangle = |\alpha_{a3}|^2 + \sinh^2(s_{31})$, and the variances of the quadrature operators change from $\Delta X_{sk1}^2 = \Delta Y_{sk1}^2 = \Delta X_{ak1}^2 = \Delta Y_{ak1}^2 = 1/4$ to $\Delta X_{sk1}^2 = \Delta Y_{ak1}^2 = \frac{1}{4}e^{-2s_{31}}$ and $\Delta Y_{sk1}^2 = \Delta X_{ak1}^2 = \frac{1}{4}e^{2s_{31}}$ (see Fig. 3). In conclusion, from the quantum theory perspective the impact of an unmatched temporal boundary is not limited to generating a reflected wave. It changes the photon statistics, generating nontrivial quantum correlations between the forward and backward waves in the form of a two-mode squeezing transformation. Antireflection temporal coatings solve this problem, not only inhibiting the generation of a backward wave, but by preserving all photon statistics while shifting the frequency of the quantum state.

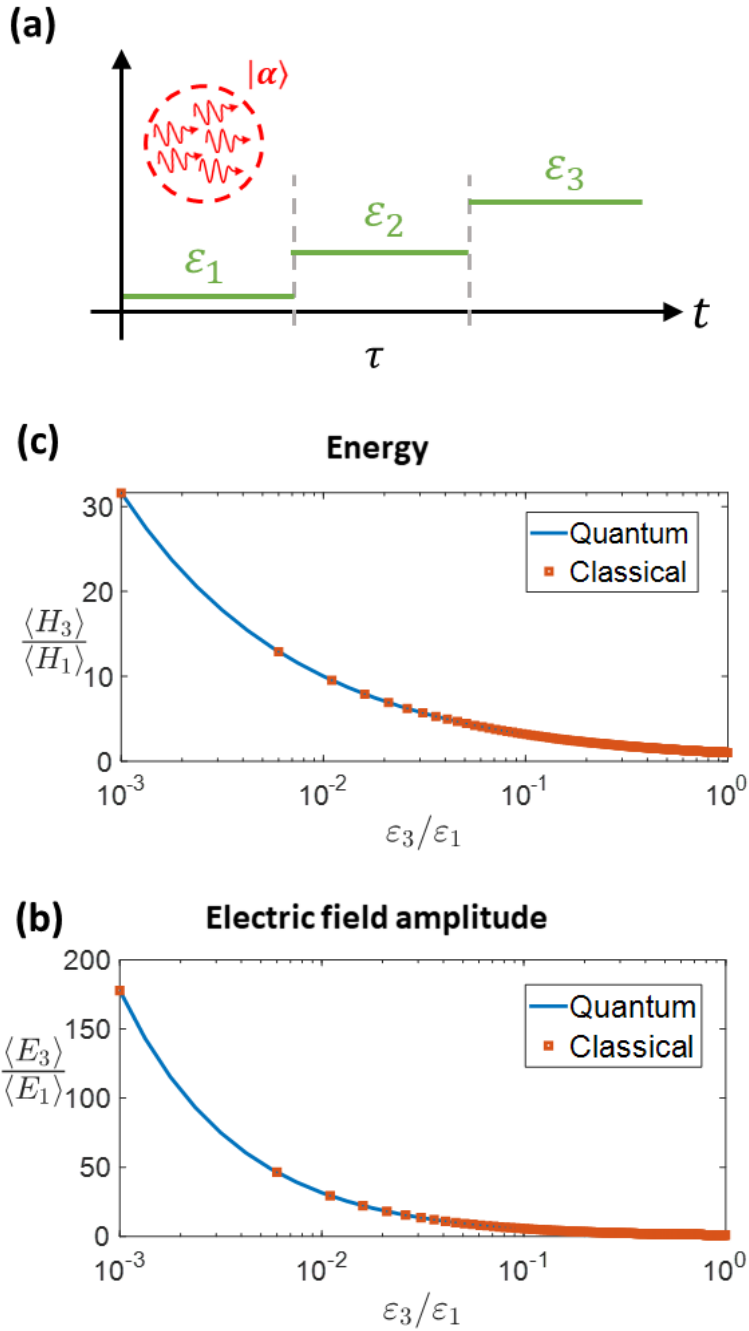

*Fig. 2. Revisiting the classical response of a quantum antireflection temporal coating.* (a) Sketch of a quantum antireflection temporal coating excited with a "classical" coherent state $|\alpha\rangle$. Change in the (b) energy and (c) electric field amplitude as a function of the permittivity contrast. Comparison with the classical prediction of the energy and electric field amplitude, calculated following [10,12], shows an excellent agreement. Note that the results for the $\varepsilon_3/\varepsilon_1 > 1$ range would simply be the inverse of these results.

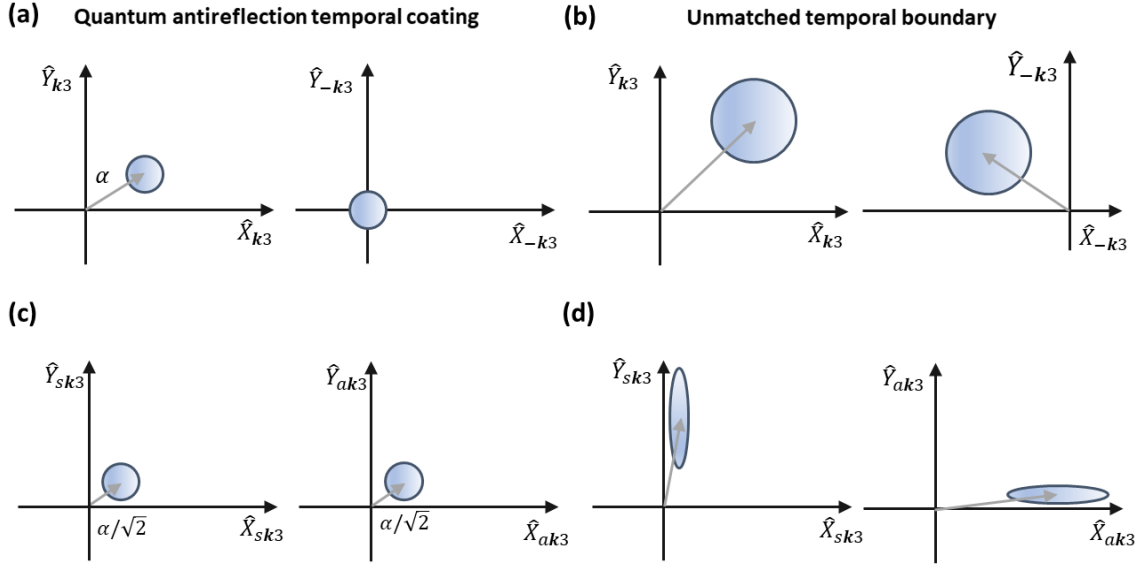

*Fig. 3. Photon statistics after the transformation of a coherent state.* Sketch of the magnitude and variance of the quadrature operators in the forward/backward (a)-(b) and symmetric/antisymmetric basis, after (c),(d) a quantum antireflection temporal coating (a),(c) and (b),(d) an unmatched temporal boundary. The figure illustrate how an unmatched temporal boundary qualitatively changes the photon statistics of a "classical wave", while a quantum antireflection temporal coating preserves the photon statistics of a coherent state.

### 4. Thermal fields and fast switching with suppressed noise

Antireflection temporal coatings can also be beneficial for avoiding the generation of noise is fast switching mechanisms. To illustrate this point, consider a material body at thermal equilibrium with a background environment at temperature $T_e$ (see Fig. 4). In such scenario, the optical modes are thermally populated, which can be described through the following density matrix [28]:

$$\hat{\rho}_{th} = \otimes_k \left(1 - e^{-\beta_{k1}}\right) \sum_{n_{k1}=0}^{\infty} e^{-\beta_{k1} n_{k1}} |n_{k1}\rangle\langle n_{k1}| \qquad (5)$$

with $\beta_{k1} = \hbar \omega_{k1}/k_B T_e$. Consequently, the average number of photons populating the modes is $\langle n_{k1} \rangle = \left(e^{\beta_{k1}} - 1\right)^{-1}$, and the variances of the quadrature operators are $\Delta X_{k1}^2 = \Delta Y_{k1}^2 = \frac{1}{4}(1 + 2\langle n_{k1}\rangle)$.

Assume that we wish to rapidly switch in time the permittivity of the medium from $\varepsilon_1$ to $\varepsilon_3$, a process that can be modelled by a single temporal boundary, with transformation rule $\hat{a}_{k3} = \cosh(s_{31})\hat{a}_{k1} - \sinh(s_{31})\hat{a}^\dagger_{-k1}$. It is expected that the interaction of the temporal boundary with the thermal fields will lead to amplification effects of the thermal noise.

Specifically, the average number of photons after the temporal boundary is $\langle n_{k3} \rangle = \langle n_{k1} \rangle \cosh^2(s_{31}) + \sinh^2(s_{31})$. The first term corresponds to the amplification of the thermal noise. The second term comes from a quantum vacuum amplification effect, akin to the dynamical Casimir effect [35,36]. We note that the number of photons always increases, independently of whether the final permittivity value is smaller $\varepsilon_3 < \varepsilon_1$ or larger $\varepsilon_3 > \varepsilon_1$ than the original one. The larger the permittivity contrast, the larger the photon production is. Similarly, the variances of the quadrature operators increase to $\Delta X_{k3}^2 = \Delta Y_{k3}^2 = \frac{1}{4}(1 + 2\sinh^2(s_{31}) + 2\langle n_{k1}\rangle \cosh^2(2s_{31}))$.

Our analysis confirms that the interaction between thermal fields and the switching mechanism entails the amplification of thermal noise. In turn, this additional noise will have a detrimental impact of the performance of the system. For example, amplifying thermal fields effectively increases the noise temperature of the system, and it will lead to photon counts on detectors located outside the material system. In addition, the amplification of the thermal fields implies that the energy of the electromagnetic system is increased. The larger the noise of the system, the larger the amount of energy that must be pumped into it, negatively affecting the energy efficiency of the switching process.

Antireflection temporal coatings, however, have the ability to mitigate the amplification of thermal noise. Since the transformation rule reduces to $\hat{a}_{k3} = \hat{a}_{k1}$, the thermal state is preserved, including the average number of photons, the variances of the quadrature operators, and any other photon statistic. Thus, antireflection temporal coatings make it possible to rapidly switch the system without amplifying the thermal noise. The cost is to reduce the speed of the process to a quarter period (at the frequency associated with mode of interest within the temporal matching layer).

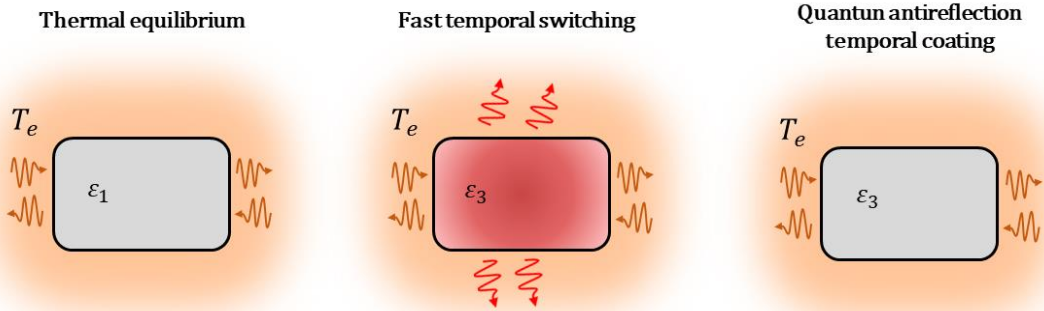

*Fig. 4. Inhibiting thermal noise amplification during fast switching. (Left) A system characterized by relative permittivity $\varepsilon_1$ being at thermal equilibrium with a background at temperature $T_e$. (Center) Amplification of thermal fields after fast switching the permittivity of the body to $\varepsilon_3$. (Right) A quantum antireflection temporal coating inhibits the amplification of thermal noise.*

## 5. Conclusions

Our results highlight the importance of investigating the quantum optical response of time-varying media. By studying the quantum response of antireflection temporal coatings, we were able to provide a clearer perspective of their classical operating principle, bringing a fresh perspective to recent studies. In addition, our formalism suggested two new potential applications of antireflection temporal coatings: quantum frequency shifting for photonic quantum networks, and fast material switches that do not amplify the thermal fields in the system. We expect that our results will motivate further research in the field, combining quantum optics and time-varying media. For example, our results could be extended to more sophisticated antireflection temporal coatings based on multilayered media, providing control over the spectral response.

## Acknowledgements

I.L. acknowledges support from Ramón y Cajal fellowship RYC2018-024123-I and ERC Starting Grant 948504. V.P-P acknowledges support from Newcastle University (Newcastle University Research Fellowship).